# Data-driven Investigation of Cotton Fabric Behavior Modified by Straight and Zig-Zag Stitches

Harmony Werth,[a#] Kazi Zihan Hossain,[a#] Momena Monwar,[a] and M. Rashed Khan[a*]

[a]Department of Chemical and Materials Engineering, University of Nevada, Reno; Reno, Nevda-89557, USA;

[#] These authors contributed equally to the work.
[*] Corresponding author. E-mail address: mrkhan@unr.edu

**Keywords**: Smart Wearables, Three-network model, Cotton Fabric, Data-informed modeling.

**Abstract:**

In this article, we demonstrate a data-driven approach to investigate the behavior of cotton fabric modified by straight and zig-zag stitches. Existing literature in understanding the mechanical behavior of soft materials (e.g., textile-based fibers or fabrics) heavily relies on stress-strain analyses. However, the strain-induced deformation behavior can be further analyzed by taking advantage of data-driven constitutive models. Such an approach reveals intermolecular parameters that can be utilized further in design and development analyses. For that, we exhibit the altered mechanics of base cotton fabric induced by two types of singular stitches (straight and zig-zag). We have sewn simple straight and zig-zag cotton stitches to investigate the mechanics of the base cotton fabrics using uniaxial stress-strain experimental data. Then, we leveraged the constitutive models (i.e., three-network model, TNM) obtained from MCalibration software to reveal eleven intermolecular parameters for data-driven investigations. Our experimental analyses, combined with the data, suggest a 99.99% confidence in assessing the mechanical impact of stitches on cotton fabrics. We have also used distributed strain energy to analyze the mechanics and failure of the base and stitched fabrics. Once adopted, our study may contribute to an improved understanding of the production of smart wearables and e-textiles.

**TOC figure:**

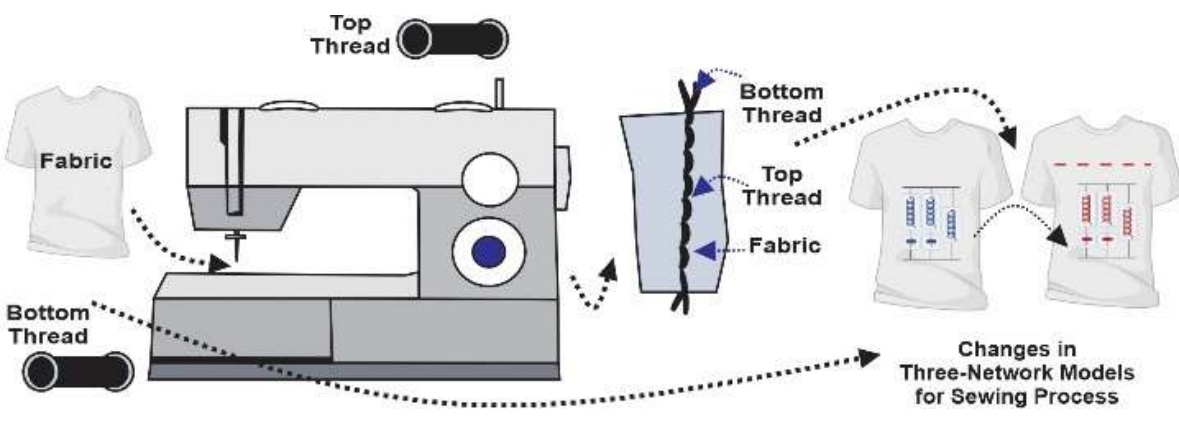

## 1. Introduction

From everyday garments to medical textiles, cotton fabrics have ubiquitous usages and have emerged as the most used soft and fibrous materials for generating and studying smart wearables. Most advanced and applied demonstrations of cotton fabrics require in-plane stitches to create new functional materials; however, stitches are randomly generated and optimized through numerous trials, leaving a fundamental gap in understanding the impact of a single sewn stitch on the fabric. Previous literature has reported the deformation-induced anisotropic mechanical behavior of the knitted cotton fabric through a stress-strain test.[1] Several studies focused on understanding knitted fabrics, fabric elongations, deformations, and failures at critical applied stress.[2–8] In contrast, the structural deformation of the fabric induced by a single stitch has remained underexplored and underutilized in the literature. Moreover, from the design of fashion to human-centered smart wearables, a data-informed approach to dissecting the role of sewn stitches in manipulating the final fabric properties has been a scientific challenge in the literature. Based on our successful preliminary results,[9] in this work, we bridge these knowledge gaps by investigating the deformation mechanics of a cotton fabric induced by a single sewing stitch using the three-network materials model (TNM).[10,11]

Our study focuses on understanding the tuned mechanics of cotton fabrics by sewn stitches and unravels data that we often overlook from simple stress-strain analyses. Generally, these stitches are randomly generated and optimized through numerous trials focusing on just aesthetic or fashion purposes, leaving a fundamental gap in how they can be optimized and functionalized further. From the material science perspective, when an alloy or composite is made, the properties of the components are bunched together to achieve an overall composite behavior.[14] The stitching or sewing process can also be studied similarly, where base fabric with added thread achieves an overall composite property.[15] Sewn stitches create entanglements between two threads- an upper and a bobbin thread (i.e., the bottom thread). During the entanglement, the sewing needle loops the upper thread between the bottom thread through the fabric, and the threads entangle. When the tension of both threads matches, the entanglement lays in the plane of the fabric on both sides with minimal damage.[12] The resulting stitch can be varied in numerous ways to create functional and non-functional patterns based on the types of sewing threads, fabric types, or choice of materials for the final composite structures.[13]

Concurrently, stitches bind two or more layers of fabric together, known as a seam. A few studies have investigated seams in woven cotton fabric to determine the impact of stitches on the mechanical behavior of fabrics.[16–18] A few investigations on the behavior of seams in knitted fabrics also exist in the current literature.[19,20] Different sewing stages,[21] sewing parameters,[22] and sewing machines[23,24] have also been reported to alter the properties of the base fabric. Alternative stitch patterns also influence the tensile and stress-relaxation behaviors of the fabric.[25,26] Prior studies also focus on understanding fracture or failure mechanisms targeting specific applications,[15,27,28] ignoring investigations at a single-stitch level, which we aim to enable in this study. Such a knowledge gap at the single stitch level motivates this work. Fig. 1 depicts the idea that we propose in this study. We hypothesize that the stitching process shifts the base fabric's strain energy density (SED) and shows a composite behavior based on the stitched materials and the base fabric. We believe studies involving a single stitch to modify the mechanical behavior of cotton fabric will become crucial for designing advanced fiber-based composite materials for future applications, offering numerous embodied intelligence and modular functionalities to perform *in situ*, biochemical, and biological analyses currently non-existent in literature.[29,30]

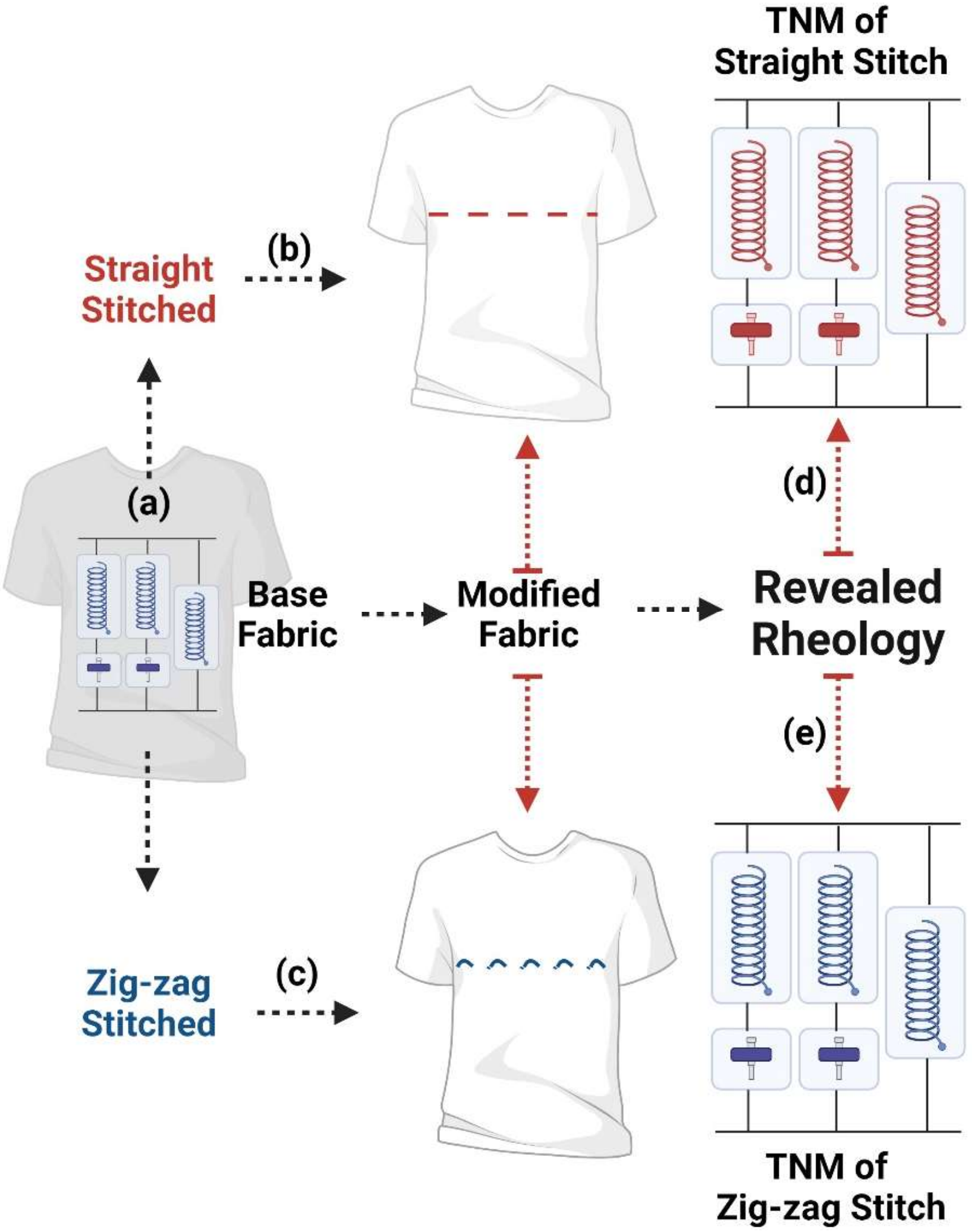

Fig. 1**: Overview of the work**. We hypothesize that (a) an unaltered base fabric can be modified via (b) a straight stitch, or (c) a zig-zag stitch, shifting the functional behavior. Corresponding TNM models will reveal the modified intermolecular properties of (d) the straight-stitched and (e) a zigzag-stitched samples.

The base fabric and the thread used in this work are made from 100% cotton- a soft matter. We assume both as an elastomeric network. Elastomers having 3,000 to 10,000 repeating units exhibit structural flexibility and experience stretch-induced softening/hardening (also known as Mullins damage) under applied loads.[31] We leverage TNM in MCalibration software[11] for a data-informed study to map the entire stress-strain spectra from the uniaxial tensile test. The TNM is also known as a constitutive material model to describe deformation-induced structural evolution (i.e., the transition between soft to stiff network) and how strain energy density (SED) becomes redistributed (i.e., hysteresis) throughout the material. Different hyperelastic materials models except TNM have been utilized in the literature to represent the mechanical behavior of fabrics.[32–36] However, we prefer TNM for its simplicity in revealing the intermolecular parameters of cotton fabric modified by a single sewing stitch. Further, TNM can be modified through MCalibration if analyses on the subsequent stitches are required for better performance. In our earlier work, we established the workflow to utilize the TNM to predict the behavior of hollow thermoplastic polymers for achieving ultra-long microfluidic channels directly on the cotton fabric.[9] This work solely focuses on the impact of the cotton fabric due to simple stitches that will enable more basic research opportunities using different types of fibrous fabrics and threads. The kinematics of the TNM consists of three parallel molecular networks where two networks are spring-dashpot domains, and the third one is only a spring capturing the large strain response of the sample. A detailed explanation of the TNM can be found in previous work.[10] Fig. 1(a) shows an illustration of the hypothetical TNM of unstitched fabric, and Fig. 1(b) and 1(c) represent the condition of the fabrics with straight and zig-zag stitching. Fig. 1(d) and 1(e) represent the respective hypothetical TNM based on altered intermolecular parameters due to two different types of stitches, which we have quantified and discussed in the latter sections.

We have chosen two types of sewing stitches, straight and zig-zag, to establish and reveal intermolecular parameters of cotton-stitch-modified cotton fabrics harnessing TNM. These stitches are common in sewn garments and are pre-programmed into the default settings of modern sewing machines. We further investigate several variations of the zig-zag stitch with varying lengths and widths. For our analyses, we investigate the surface topography of the fabric and samples with sewn stitches using optical and scanning electron microscope (SEM) images. We perform (a) uniaxial stress-strain and (b) repeated cyclic tests in a tensile testing instrument to dissect the tensile behaviors of the (a) base fabric, (b) base threads, and (c) threads-laid-fabric structures. The uniaxial stress-strain analyses have revealed three regions of interest (i.e., elastic, yield, and viscoplastic). Also, we have investigated the permanent failure of the entire composite (fracture) to find the extremities of experimental analyses. The cyclic tests provide information about hysteresis, which we leverage to understand distributed SED and the loss due to hysteresis. We outline calibration using TNM in MCalibration to offer a simple route to test the impact of specific sewing patterns on the intermolecular behavior of the final stitched fabric. Our experimental analyses, combined with the calibrated data, suggest a 99.99% confidence in assessing the influence of a single stitch on knitted cotton fabrics. Our data-driven study may enable the fundamental investigations related to the optimum design and study of integrating smart threads in cotton fabrics to produce smart wearables, e-textiles, and biomedical and e-fashion textiles for the future.

## 2. Materials and Methods

We analyzed uniaxial stress-strain curves for each component (cotton fabric and cotton thread) and different variations of the overall composites (straight stitched fabric, zig-zag stitch). We further obtained their materials model, TNM, and compared the properties of their models.

### 2.1 Materials

The fabric used in the experiments was 100% cotton jersey knit with a unit weight of 427 g/m$^2$ obtained from Hobby Lobby Stores, Inc. The measured thickness of the fabric was 0.45 mm. Similarly, the thread used in this experiment was a 50-weight, 4-ply, 100% cotton thread of Sew-Ology Brand from Hobby Lobby Stores, Inc., produced for traditional stitching. The outer diameter of the thread measured with a Stereoscope was 0.30 mm. The same thread was used for the top and bobbin threads for all samples prepared and presented in this work.

### 2.2 Sample Preparation

We used a Brother SE600 sewing machine bought from Amazon to create stitches of manually adjustable length and width. The tension was selected to maintain equal tension on the bobbin thread and the upper thread, preventing the bottom thread from showing on the top or vice-versa, which is a common problem during sewing practice. A swatch of fabric approximately 20 cm long was cut with scissors and then sewn wale-wise with the appropriate type of stitch for the sample. Two samples were prepared with stitching: straight stitched and zig-zag stitched. The straight stitch was 2 mm in length. The zig-zag stitches were (listed as stitch length x stitch width) 2x5 mm, 1x5 mm, 3x5 mm, 2x3 mm, and 2x4 mm. Several samples without any sewn stitches were also prepared for comparison. The fabric samples were then cut to the same size with a Cricut Maker fabric cutter bought from Amazon, allowing for accurate and reproducible sample cutting. For every sample, the fabric was cut to 6 cm in length by 2 cm in width. Sample cutting was done carefully to keep the stitching in the center of the sample.

Damaged or samples with uncentered stitches were discarded without any analysis.

### 2.3 Image Acquisition

Olympus SZ61 stereomicroscope paired with an Amscope MU1000-HS camera was used to capture the top-down optical microscopic images. Scanning electron images (SEM) of the samples were analyzed using a Thermo Scientific Scios 2 SEM. Small representative samples were cut and loaded on the holder with double-sided carbon tape for SEM imaging. Attention was given to keeping the stitch undamaged while loading on the holder. Since the samples were nonconductive, they were sputter-coated with gold (Au) to create a ~10 nm layer on their surface to avoid the charging effect (bright spots) before imaging. Further optical images were captured with the camera on an iPhone 13 mini.

### 2.4 Experimental Methods

The data was collected using an Instron 5982 for uniaxial tensile testing within a 4cm x 2cm area. The extra centimeters on each side of the samples allowed the grip to hold them during testing. Each sample type was examined with tensile testing to determine the stress and strain until failure at a 40 mm/min strain rate. Cyclic testing of four cycles was then conducted for specific samples up to a sustainable strain level for that sample type. Samples with straight stitches could only withstand slightly more than 10% strain. Therefore, cyclic tests with straight stitches were conducted up to 10% strain. For comparison, cyclic testing up to 10% strain was also conducted for unaltered fabric samples and the 2x5 mm zig-zag stitch.

### 2.5 Strain-energy Density Calculation

In this work, we used two similar terms: strain energy density (SED) and strain or stored energy (SE). The SED of the samples was calculated using the trapezoidal rules to integrate the area under the stress-strain curve. The curve was divided into equal-time steps, and these small areas were added till the last point to get the total area under the curve.[9] All the SED was calculated up to 4.5% strain for comparison because the thread samples failed around 5% strain. Table S1 includes a detailed SED comparison between the samples investigated here.

SE was calculated by multiplying the volume of the samples with the experimentally measured SED. For the stitched samples, we compared two types of SE— measured and predicted. Measured values were obtained directly from the tensile behavior of the stitched samples, i.e.,

$$Measured\ SE_{Stitched\ Sample} = (Volume_{Fabric} + 2\ X\ Volume_{\ Threa}\ )X\ Measured\ SED_{Stitched\ Sample}$$

On the other hand, the predicted values were obtained by adding the individual SE of the components of the stitched samples, i.e.,

$$Predicted\ SE_{Stitched\ Sample} = Volume_{Fabric}\ X\ Measured\ SED_{Fabric} +$$

$$2\ X\ (Volume_{\ Threa}\ \ X\ Measured\ SED_{\ Threa}\ )$$

For both equations, thread properties were multiplied by 2, as a stitched composite consists of the top and bobbin threads.

### 2.6 Constitutive Modeling of Different Composites

MCalibration, from PolymerFEM,[11] was used to obtain parameters for a material model capable of representing the intermolecular behavior of the fabric samples prepared in this work. MCalibration fits the experimentally collected uniaxial stress-strain data to the PolyUMod Three Network model to obtain the representative intermolecular model.[9] An average of the stress-strain behavior with the time of each sample type was obtained first from multiple experimental runs at the same strain rate. The averaged data were then processed using the MCalibration software tools to prepare the data for calibration. MCalibration software begins calibration with initially estimated parameter values by observing the experimental data. It tries to reduce the deviation between the predicted and the observed behavior by continuously updating the parameters. The default settings were used to achieve the calibration. The calibrated parameters were exported and analyzed after the automatic convergence of the calibration process.

## 3. Results and Discussion

### 3.1 Surface Topography

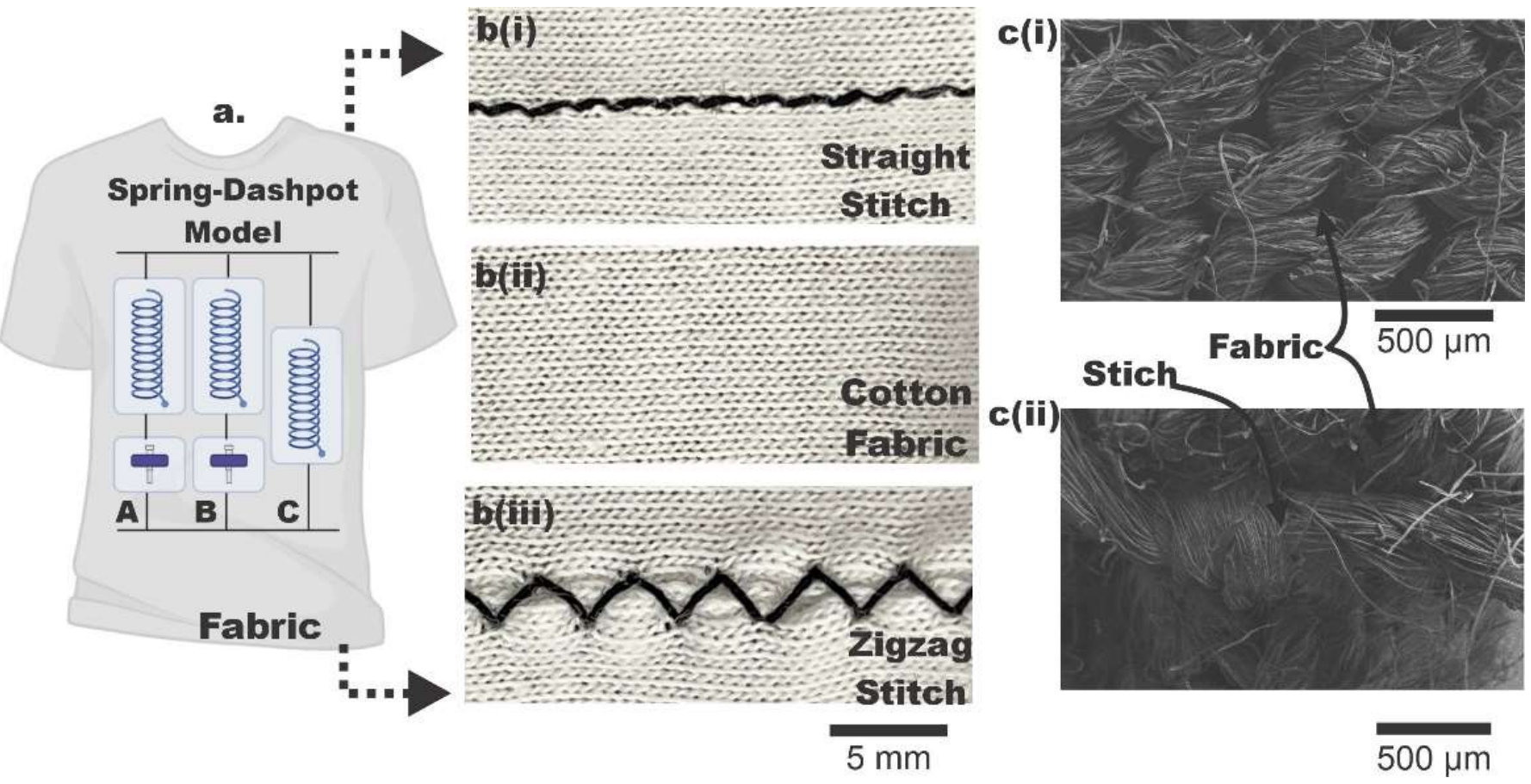

Fig. 2: **Topographic observations**. A cartoon (a) representation of the Three-Network spring-dashpot model for tensile behavior of cotton fabric. We inspected all samples under optical microscope in b(i-iii), and acquired (c) surface topography of (i) a bare cotton fabric and (ii) sewn stitched-cotton fabric under SEM. Scale bars are shown at the bottom.

We formed two types of stitches on the base fabric for topographical observations, as shown in Fig. 2. Fig. 2(a) is the graphical depiction of the three-network model, consisting of spring and dashpot used in this work. Fig. 2(b) are top-down optical microscope images of the base and sewn fabrics for visual inspections and to identify the differences between a straight stitch and a zig-zag stitch on the fabric. These images show that the straight and zig-zag stitches went through the fabric without significant internal damage. The straight stitch shown in Fig. 2b(i) does not have substantial bunching due to the stitch compared with the only fabric shown in Fig. 2b(ii); however, a meandering network of the zig-zag stitches pulled the fabric within the stitch, as shown in Fig. 2b(iii). The fabric cannot maintain its original shape due to zig-zag stitching and bunching into the stitch instead. The structural stiffness and flexibility of the fabric may have contributed to the bunching, as observed within the stitch dimensions in Fig. 2(b)(i-iii). Fig. 2(c) are the SEM images of the sewn fabric to investigate the surface topography of the stitches and fabric in higher magnification. SEM images in Fig. 2c(i) reveal the undamaged fabric at the thread level, and Fig. 2c(ii) shows the knot on the fabric by the sewing thread. We performed the uniaxial tensile tests to understand the altered intermolecular behavior of the bare fabric and stitched composites.

**3.2 Uniaxial Tensile Behavior**

Using Instron 5982 for tensile behavior analyses, we investigated plain thread, plain fabric, and stitched fabrics. The uniaxial tensile test behavior of cotton thread is shown in Fig. 3(a), and the fabric is shown in Fig. 3(b). For comparison, Fig. 3(b) also shows the behaviors of straight and zig-zag (2x5mm) stitched fabrics. We compared four zig-zag stitches varying in length or width on cotton fabrics in Fig. 3(c) to understand the intermolecular modifications more. Fig. 3(d-e) shows the side view of a representative zig-zag stitched fabric loaded into the instrument during tensile testing and at the end of failure analyses.

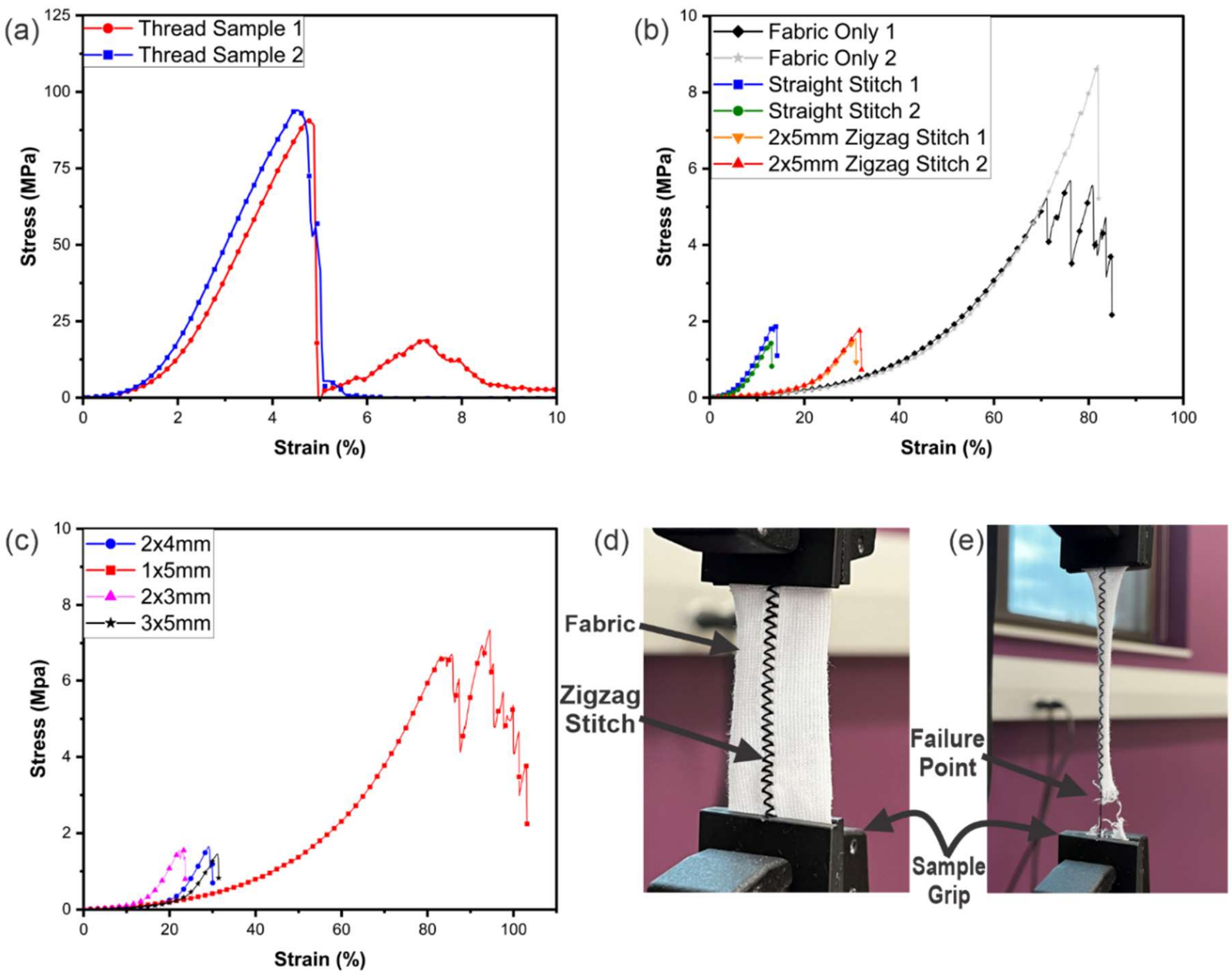

Fig. 3: **Tensile tests**. (a) Stress-strain curve for the samples of the cotton thread indicating maximum stress of approximately 100MPa at a strain of ~4.5% before failure, (b) stress-strain curves of the unaltered fabric, fabric with straight stitches and fabric with zig-zag stitches where stitch length×width is 2mm×5mm, (c) stress-strain graph reflecting the impact of zig-zag stitches property variation, (d) image of a 1x5mm sample with zig-zag stitches showing the condition of the sample before uniaxial tensile loading, (e) image of a 1x5mm sample with zig-zag stitches revealing the condition of the sample after uniaxial tensile loading; Notably, the fabric has failed while the sewn thread is intact.

We tested two samples of the plain threads, and the behaviors are shown in Fig. 3(a). The plain cotton thread failed at 5% strain but exhibited the highest stress level, ~100 MPa, compared to other samples tested. In contrast to the plain thread, the cotton fabric in Fig. 3(b) exhibited reproducible stretchability of up to 70% strain in two samples. Including straight stitches into the plain fabric induced failure at ~12% strain, and the zig-zag stitched sample failed at ~32% strain. Due to the stitching process, the composite showed significant deviation from the reported alteration between 10-20% of the in-plane tensile properties. However, none of the previous works directly focused on the SED or the maximum failure criteria. Instead, the focus was on simple intermolecular properties like Young's modulus, tensile strength, or compressive strength.[27,28]

The unaltered fabric survived the highest strain at the point of failure, shown in Fig. 3(b), between 60-80%, with a SED of around 1.0 $MJ/m^3$ at failure. In contrast, the thread samples had a SED of approximately 5.0 $MJ/m^3$ at failure, which occurred at around only 5% strain. The SED of the unaltered fabric and thread at 4.5% were $4.19x10^{-4}$ $MJ/m^3$ and 4.64 $MJ/m^3$, respectively. Examining the stress-strain data for the cotton fabric and the cotton thread individually, we conclude that compositing these materials would result in a sample with SED that falls between the materials at a given strain. The samples with sewn straight stitches of 2mm length failed between 10-15% strain. At a strain of 4.5%, the straight stitched samples exhibited an average SED of $1.88x10^{-3}$ $MJ/m^3$, a ~4-fold higher SED compared with the unaltered fabric, which indicates gaining structural strength due to the composition of the thread. As the straight-stitched sample survived more than 10% strain, we added the SE of the thread until fracture and the fabric's SE until 10% and divided it by the total volume; we get the prediction value of SED $7.43x10^{-2}$ $MJ/m^3$, which is slightly different from the measured SED of the straight-stitch sample. This discrepancy is expected as the sewing process exposes the thread to dynamic loads and friction, which is known to reduce the strength of the thread.[37] Overall, the sample with sewn straight stitches failed at the lowest stress and strain compared with other fabric samples. The cause of the low stress and strain at failure is suspected to be related to the structure and orientation of the stitches or the added cotton thread, which cannot withstand as much strain as the fabric. The fabric, with a higher elongation at failure than the thread, can deform under the load. Therefore, the thread in the straight stitch withstands the load for the entire sample until the thread breaks, equivalent to sample failure. It was observed that the thread failed before the fabric in all samples with straight stitches.

Samples with zig-zag stitches of 2mm length and 5mm width also failed at stress and strain lower than the unaltered fabric but higher strain than the straight stitched sample. An analysis of the SED of the 2x5mm zig-zag sample reveals aspects of the intermolecular behavior. At strains below 20%, the SED of the 2x5mm zig-zag sample is indistinguishable from the SED of the fabric; Therefore, the fabric's intermolecular properties dominate the thread's properties in the 2x5mm zig-zag sample at strains under 20%. At 30% strain, the SED of the zig-zag sample is nearly double that of the fabric sample. The departure of the 2x5mm zig-zag sample from the intermolecular behavior of the fabric indicates that at strains above 20%, the thread is the dominant influence on the intermolecular behavior. This behavior is investigated further in zig-zag samples with varying stitch lengths and widths, as indicated in Fig. 3(c).

During uniaxial tensile testing, it was revealed that stitch length and width are critical factors that influence the tensile behavior of the samples with zig-zag stitches. Fig. 3 (c) shows stress-strain curves for samples with varying lengths and widths of zig-zag stitches. The fabric samples with 2x5mm, 2x4mm, and 3x5mm zig-zag stitches failed at a higher strain than those with straight stitches but at a similar stress. As with the 2x5mm sample, the 2x4mm and 3x5mm had similar SEDs at low strain until the thread became a dominant influence. The 1x5mm zig-zag sample exhibited drastically different behavior from the other samples with zig-zag stitches.

The SED of the 1x5mm zig-zag samples matched that of the unaltered fabric sample up to approximately 60% strain, indicating that the stitches had little impact on the tensile behavior of the sample overall. The structure of the zig-zag stitch contributes to the behavior of all the zig-zag samples. However, a stitch length of less than 2 mm seems to be unable to alter the failure criteria of the composite. This phenomenon seems reasonable as more than enough stitched thread on the perpendicular direction of the tensile testing will not significantly contribute to the failure criteria of the composite. Since zig-zag stitches have both a stitch length and a stitch width, the stitch could change shape as the fabric elongates.

Fig. 3(d) and (e) show a 1x5mm zig-zag sample before and after uniaxial tensile testing. After testing, the stitches are longer in the direction parallel to loading and shorter in the direction perpendicular to loading compared to before tensile testing. In other words, the stitch could shrink in the direction perpendicular to loading while elongating in the direction of loading. A shorter stitch length results in more threads in the sample, which allows the stitches to deform enough to match the elongation of the fabric. The consequence of the stitch deformation is that the fabric withstands the load while the stitch can deform, but the stitch bears the load when it can no longer match the elongation of the fabric. Eventually, the load exhausts the ability of the stitch to deform, which is when the SED of the zig-zag sample deviates from that of the unaltered fabric. It was observed that the sewn thread had snapped in all samples after the sample had failed during tensile testing, except for the 1x5mm zig-zag sample. The 1x5mm zigzag sample in Fig. 3(c) had a shorter stitch length. Another point of interest is shown in Fig. 3(e), which shows that the thread was intact after the fabric failed, which is the opposite of all other samples that contained sewn stitches. Therefore, it is possible to alter stitch properties to alter the fabric's tensile behavior, and the properties determine the extent of the influence from the thread and the fabric at particular strains.

Furthermore, we investigated the tensile behavior of the unaltered fabric, straight stitch, and 2x5mm zig-zag stitch samples under cyclic loading to analyze stress softening and hysteresis. All samples were strained up to 10% because the straight stitch samples failed at approximately 12% strain, as shown in Fig. 3(b). Repeated cycles allow for an investigation of the hysteresis in additional cycles and an analysis of the stress-softening behavior of the samples. Hysteresis, the change in behavior from the loading to the unloading cycle, was observed in all samples, as shown in Fig. S1 in the supporting section. The relationship between the yarns, the fabric structure, and the sewing type impacts the hysteresis. The plastic deformation of the yarns, which relates to the slippage and viscoelasticity of the fibers within the yarn, influences hysteresis.[8] The structure dictates the number and nature of the contact points between thread loops, which impacts friction during loading. Friction can also be the main factor determining the amount of hysteresis that will occur.[3] Across all samples, the most extensive hysteresis occurred during the first cycle. The second cycle revealed that stress softening occurred in all samples between the first and second cycles, which can be observed throughout Fig. S1 as a reduction in the SED of the loading curve between the first and second cycles. Additionally, all samples had the highest SED during the loading of the first cycle.

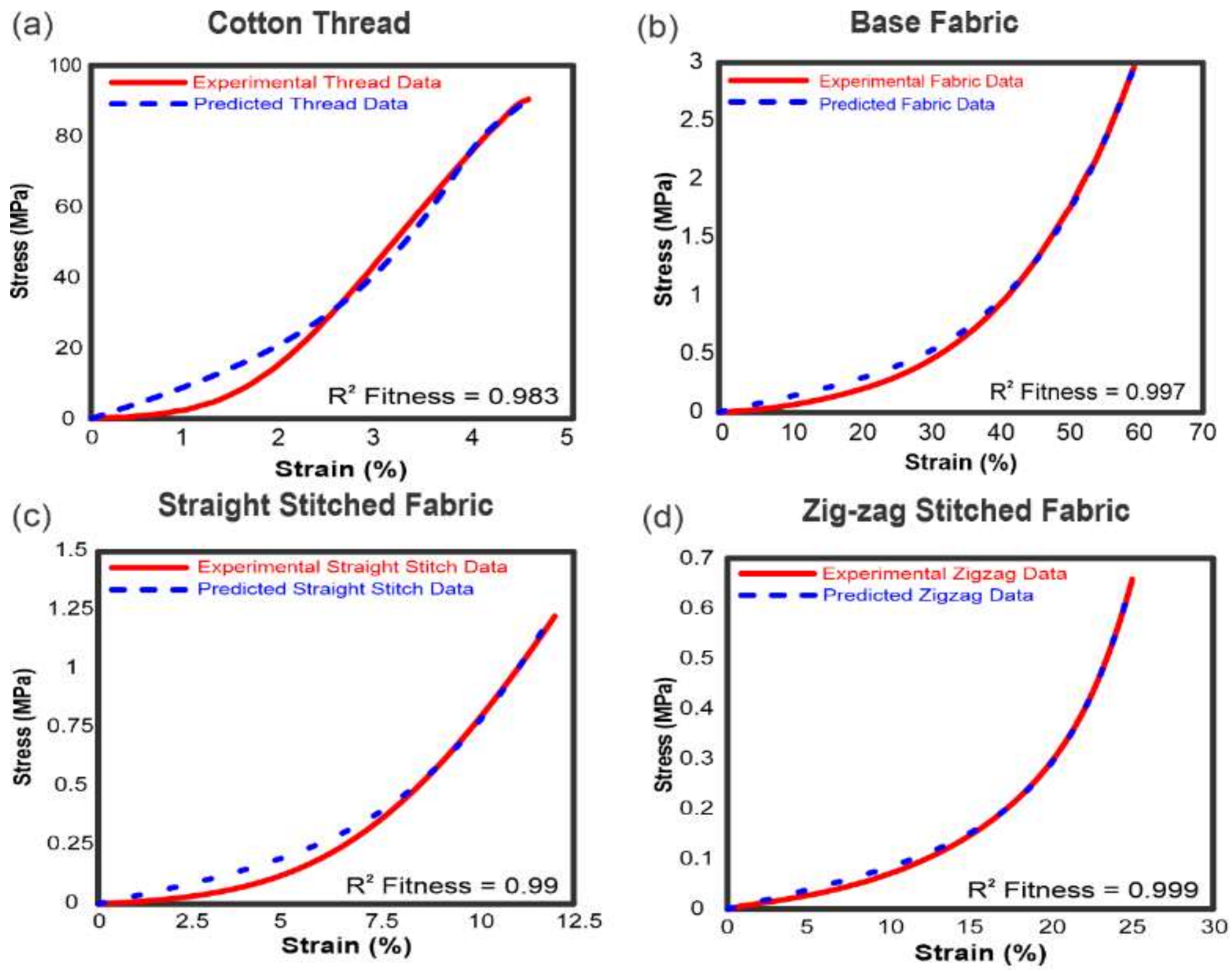

Fig. 4: **The material calibration** with the PolyUMod TNM resulted in more than 99% correlation with the experimental observation for (a) a 100% cotton thread, (b) an unaltered fabric sample, (c) the straight stitch sample, and (d) the zig-zag sample**.**

### 3.3 Revealing intermolecular parameters of Fabric and Composite Systems

We used MCalibration software to perform intermolecular analyses using TNM and calibrate the TNM parameters to assess unaltered and altered fabrics. Bergstrom and Bischoff explained the mathematical details of the TNM in their work.[10] While the stress-strain analysis directly represents the tensile behavior, the detailed intermolecular parameters are often overlooked, which can be revealed through constitutive models. When the coefficient of determination, the $R^2$ value, or the differences between the experimental and the calibrated data, stops changing significantly by reaching the convergence, the software MCalibration reveals the intermolecular parameters in its user interface. The experimental data and MCalibration predicted data with their respective $R^2$ fitness are shown in Fig. 4, indicating that the TNM model can effectively capture the uniaxial tensile behavior of unaltered cotton thread, unaltered fabric, and their composites- straight- and zigzag-stitched fabrics. The predicted data fits closely (more than an $R^2$ fitness of 0.99) with the experimental data for all investigated samples with this method. Table 1 shows several such parameters of different fabric samples.

**Table 1:** The Strain energy density (SED) and revealed Three-Network Model (TNM) parameters of cotton thread, unaltered fabric, straight stitch, and the zig-zag stitch

| Description | Unit | 100% Cotton Thread | Unaltered Fabric | Straight Stitch | Zig-zag Stitch |
| --- | --- | --- | --- | --- | --- |
| Measured SED up to 4.5% | $MJ/m^3$ | 4.64 | 4.19 x$10^{-4}$ | 1.80 x$10^{-3}$ | 5.96 x$10^{-4}$ |
| Measured SED up to the mentioned strain | | 4.64 | 1.01 (70%) | 2.35 x$10^{-2}$ (10 %) | 9.34 x$10^{-2}$ (30 %) |
| Shear modulus of network A | KPa | 29077.10 | 11.40 | 77.95 | 0.65 |
| Locking stretch | - | 1.01 | 1.08 | 1.02 | 1.04 |
| Bulk modulus | KPa | 307263.19 | 656.37 | 1369.34 | 1194.72 |
| Flow resistance of network A | KPa | 48414.97 | 127.80 | 901.91 | 352.31 |
| Stress exponential of network A | - | 9.48 | 3.83 | 11.11 | 9.59 |
| Initial shear modulus of network B | KPa | 305204.54 | 96.46 | 15.05 | 40.75 |
| Final shear modulus of network B | KPa | 14995.52 | 96.46 | 9.31 | 58.99 |
| Evolution rate of $\mu_B$ | - | 11.04 | 9.69 | 10.20 | 10.50 |
| Flow resistance of network B | KPa | 85306.96 | 348.78 | 1226.80 | 636.76 |
| Stress exponential of network B | - | 12.89 | 7.89 | 9.65 | 10.85 |
| Shear modulus of network C | KPa | 587.31 | 398.95 | 1180.98 | 207.47 |

Most of the modulus parameters of the cotton thread are $10^4$-fold higher than modulus values, as its SED also had a similar difference. In the TNM, networks A and B utilize separate energy activation mechanisms to represent polymer networks' amorphous and semi-crystalline domains. Network C represents the large strain response controlled by entropic resistance. The shear modulus and the flow resistance of network A in the straight stitch are significantly higher than the other two samples, indicating higher resistance by the spring represented in the network.

Comparatively close initial and final shear modulus of network B and almost similar evolution rates indicate a similar effective shear modulus for all the samples. The flow resistance of network B and the shear modulus of network C of the straight-stitched sample are also higher, indicating higher stiffness of the materials. We also compared the TNM parameters with previously reported hyperelastic material models having a shear modulus of 3.89 KPa and a limiting locking stretch $(\lambda_{L,lim})$ of 0.65907.[32] The shear modulus of the Network A of our unaltered fabric is also in the same range. To summarize from Table 1, we demonstrate the feasibility of studying the intermolecular mechanics and deformation of base fabric modified by two different types of singular stitches. While we have kept this study in one stitch, our study provides a foundational framework to design fabric-based composites using data-driven approaches.

**Conclusions**

This work determined that altering the stitching parameters when sewing with cotton thread into a single layer of jersey-knit cotton fabric impacts the sample's strain-energy density, hysteresis, and stress softening. When examined with optical and scanning electron microscopes, the stitched samples did not show significant damage to the fabric from the sewing process. A zig-zag stitch's stitch type and parameters were shown to impact the sample's behavior under uniaxial tensile loading directly. Depending on the stitch type, the fabric can be altered to have a higher or lower SED at certain strains. We also note that stitches capable of less elongation than the fabric will increase the SED at lower strains and result in failure at a lower strain. Stitches that can match or exceed the elongation may have minimal impact on the SED of the sample at the same strains as a sample without stitches but will fail at higher strains, resulting in a higher SED at failure. Stitches will also impact the hysteresis and stress softening of the sample. Also, stitches capable of less elongation than the fabric will be subjected to higher stress during loading, resulting in plastic deformation and more significant hysteresis and stress softening during cyclic loading. The tensile and cyclic tests reveal that the intermolecular behavior of samples composed of fabric with stitches varies greatly depending on the relationships between the property of the materials and their structure. When data from tensile tests were calibrated with the PolyUMod TNM, the materials presented in this work matched well with the calibrated model; therefore, materials calibration provides an opportunity to aid the selection of materials and structure by offering insight into hidden parameters that allow for a data-driven approach to design.

Limitations of this work include the number of materials and structures investigated, as the behavior observed may differ from samples with different compositions and structures. Furthermore, sewing stitches not examined in this paper may impact many other properties, such as abrasive strength, bursting strength, torsional properties, ability to withstand washing and drying, and many other characteristics. Future works may investigate the impact of additional types of stitches on fabrics of different materials and structures and analyze additional properties of the samples.

**Conflicts of interest:**

The authors declare no conflicts of interest.

**Acknowledgements**

MRK acknowledges the funding support from VPRI's startup fund. HW acknowledges the Nevada Undergraduate Research Award (NURA) fund from the Undergraduate Research Office, and KZH acknowledges funding from the College of Engineering Dean's Office at the University of Nevada, Reno. HW acknowledges contributions from Sydney Fields, Jake Kattelman, Thomas Kaps, and Braden Norris for the MSE 470 (Polymer Engineering instructed by MRK) in-class project. KZH acknowledges the opportunity to train and mentor all the groups in CHE/MSE 470 and Brian Perdue in CHE 495 to test and validate the concepts of this article.

## Appendix A. Supplementary data

Supplementary data related to this article can be found at Journal Website. All the relevant strain-energy densities have been compared here.

Table S1: Strain Energy density (SED) comparison between different Sample

| Sample Name | | Average Total SED upto mentioned Strain Level, before breakdown (MJ/$m^3$) | SED at 4.5 % strain level (MJ/$m^3$) |
|---|---|---|---|
| Cotton Thread | | 4.64 (4.5 %) | 4.64 |
| Cotton Fabric | | 1.01 (70 %) | 4.19 x$10^{-4}$ |
| Cotton Fabric with Straight-stitch | | 2.35 x$10^{-2}$ (10 %) | 1.80 x$10^{-3}$ |
| Cotton Fabric with Zigzag-stich | 2x4mm | 4.08 x$10^{-2}$ (20 %) | 4.22 x$10^{-4}$ |
| | 2x5mm | 9.34 x$10^{-2}$ (30 %) | 5.96 x$10^{-4}$ |
| | 1x5mm | 1.13 (80%) | 6.22 x$10^{-4}$ |
| | 2x3mm | 4.39 x$10^{-2}$ (20 %) | 5.98 x$10^{-4}$ |
| | 3x5mm | 1.24 x$10^{-2}$ (20 %) | 3.45 x$10^{-4}$ |

## Repeated Cycling Behavior

The unaltered fabric sample in Fig. S1(a) showed a minor stress softening, which can be attributed to the significant difference between the maximum strain during cyclic loading and the strain required to cause failure. Since the unaltered fabric sample has minor unrecoverable deformation at the 10% strain tested in this experiment, minimal stress softening occurred. In additional cycles after the second cycle, hysteresis in the fabric and zig-zag samples remained the same; however, hysteresis decreased slightly in the straight stitched sample from the second to the third cycle. The decrease in hysteresis is attributable to the stress softening in the straight stitch sample between the second and the third cycles, which indicates that further unrecoverable deformation occurred during each cycle. The fabric and the zig-zag stitch samples do not experience significant unrecoverable deformation in cycles after the second cycle. In the stiched samples, the causes of tensile hysteresis are further complicated by the presence of the stitched threads, which impact the overall properties and structure of the sample. In Fig. S1(b), the straight stitch sample showed more hysteresis than the zig-zag stitched sample shown in Fig. S1(c), indicating that the straight stitched threads experienced more plastic deformation than the zig-zag stitched threads. The straight-stitched thread sustains significantly more stress for the sample than the normal fabric and the zig-zag stitch. The maximum stress level at a similar strain is ~16 times higher for the straight-stitched sample. This indicated that the sewing thread in the straight-stitched sample is a remarkably dominant factor dictating this composite sample's cyclic behavior and showed more plastic deformation.

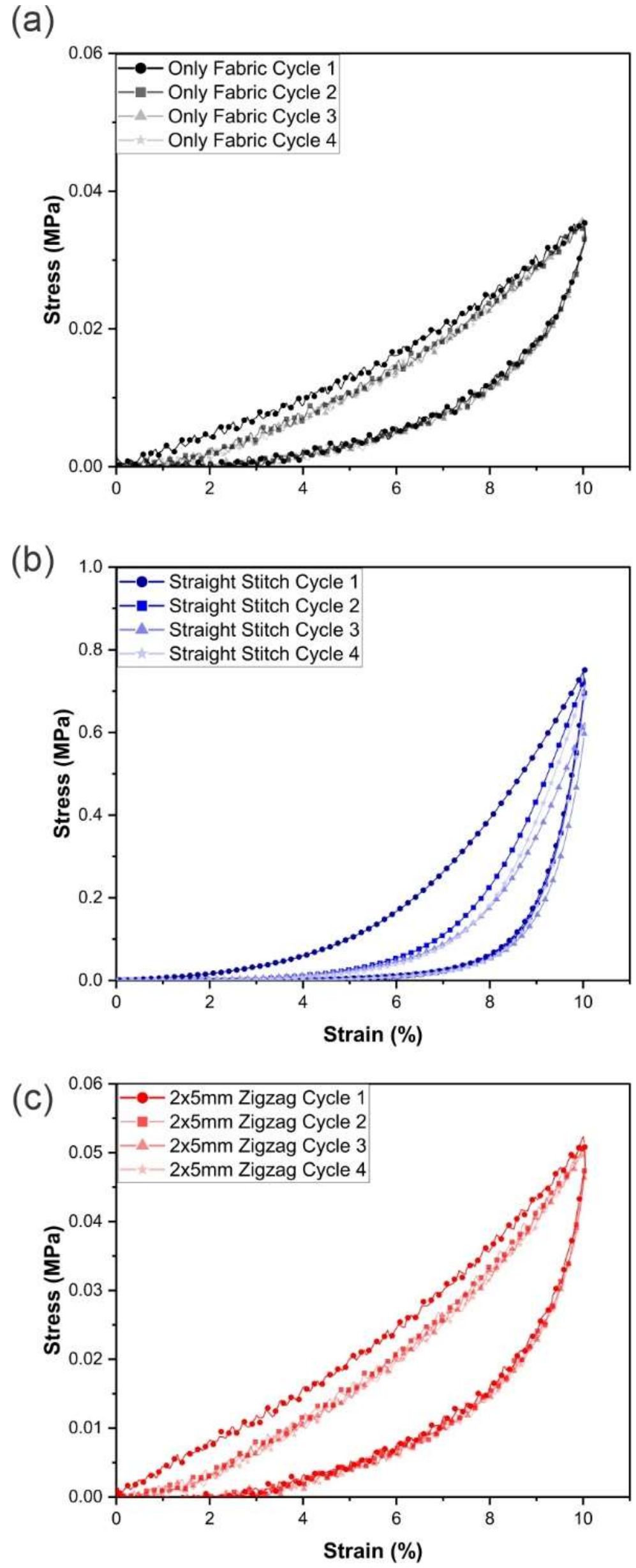

Fig. S1: Hysteresis **plot** for cyclic loading of (a) an unaltered fabric sample, (b) a sample with straight stitch, (c) a sample with zig-zag stitch. The straight stitch sample had the most stress softening compared to the others. The zig-zag stitched sample showed stress softening after the first cycle, where the sample size was 2×5mm. Also, the unaltered cotton sample showed less stress softening than the zig-zag stitched sample. Here, the x-axis value level is same for all (a)-(c). However, y-axis level in (b) is significantly different compared with (a) and (c).

**References:**

(1) Penava, Ž.; Penava, D. Š.; Miloš, L. Experimental and Analytical Analyses of the Knitted Fabric Off-Axes Tensile Test. *Textile Research Journal* **2021**, *91* (1–2), 62–72. https://doi.org/10.1177/0040517520933701.

(2) Mohamed, A.; Messiry, M. E. Analysis Of The Effect Of Cyclic Loading On Cotton-Spandex Knitted Fabric; 2016; pp 1–6. https://doi.org/10.15224/978-1-63248-090-3-41.

(3) Dusserre, G. Modelling the Hysteretic Wale-Wise Stretching Behaviour of Technical Plain Knits. *European Journal of Mechanics - A/Solids* **2015**, *51*, 160–171. https://doi.org/10.1016/j.euromechsol.2014.12.009.

(4) Li, Q.; Wang, Y.; Jiang, S.; Li, T.; Ding, X.; Tao, X.; Wang, X. Investigation into Tensile Hysteresis of Polyurethane-Containing Textile Substrates for Coated Strain Sensors. *Materials & Design* **2020**, *188*, 108451. https://doi.org/10.1016/j.matdes.2019.108451.

(5) Andrews, B. A. K.; McSherry, W. F.; Frick, J. G.; Cooper, A. B. Recovery from Tensile Strain in Knitted Cotton Fabric after Cross-Linking. *Textile Research Journal* **1971**, *41* (5), 387–391. https://doi.org/10.1177/004051757104100503.

(6) Choi, M.-S.; Ashdown, S. P. Effect of Changes in Knit Structure and Density on the Mechanical and Hand Properties of Weft-Knitted Fabrics for Outerwear. *Textile Research Journal* **2000**, *70* (12), 1033–1045. https://doi.org/10.1177/004051750007001201.

(7) Liu, R.; Lao, T. T.; Wang, S. X. Impact of Weft Laid-in Structural Knitting Design on Fabric Tension Behavior and Interfacial Pressure Performance of Circular Knits. *Journal of Engineered Fibers and Fabrics* **2013**, *8* (4). https://doi.org/10.1177/155892501300800404.

(8) Abdessalem, S. B.; Abdelkader, Y. B.; Mokhtar, S.; Elmarzougui, S. Influence of Elastane Consumption on Plated Plain Knitted Fabric Characteristics. *Journal of Engineered Fibers and Fabrics* **2009**, *4* (4). https://doi.org/10.1177/155892500900400411.

(9) Hossain, K. Z.; Khan, M. R. Computer-Aided Engineering of Stretchable Hollow Fibers to Enable Wearable Microfluidics. *Advanced Materials Technologies* **2023**, *8* (18), 2300682. https://doi.org/10.1002/admt.202300682.

(10) Bergstrom, J. S.; Bischoff, J. E. An Advanced Thermomechanical Constitutive Model for UHMWPE. *The International Journal of Structural Changes in Solids* **2010**, *2* (1), 31–39.

(11) Bergstrom, J. S. *Mechanics of Solid Polymers: Theory and Computational Modeling*; William Andrew, 2015.

(12) Ukponmwan, J. O.; Mukhopadhyay, A.; Chatterjee, K. N. Sewing Threads. *Textile Progress* **2000**, *30* (3–4), 1–91. https://doi.org/10.1080/00405160008688888.

(13) Yıldız, E. Z.; Pamuk, O. The Parameters Affecting Seam Quality: A Comprehensive Review. *Research Journal of Textile and Apparel* **2021**, *25* (4), 309–329. https://doi.org/10.1108/RJTA-05-2020-0044.

(14) Clyne, T. W.; Hull, D. *An Introduction to Composite Materials*; Cambridge university press, 2019.

(15) Song, C.; Fan, W.; Liu, T.; Wang, S.; Song, W.; Gao, X. A Review on Three-Dimensional Stitched Composites and Their Research Perspectives. *Composites Part A: Applied Science and Manufacturing* **2022**, *153*, 106730.

https://doi.org/10.1016/j.compositesa.2021.106730.

(16) Sülar, V.; Meşegül, C.; Kefsiz, H.; Seki, Y. A Comparative Study on Seam Performance of Cotton and Polyester Woven Fabrics. *The Journal of The Textile Institute* **2015**, *106* (1), 19–30. https://doi.org/10.1080/00405000.2014.899079.

(17) Rogina-Car, B.; Schwarz, I.; Kovačević, S. Analysis of Woven Fabric at the Place of the Sewn Seam. *AUTEX Research Journal* **2018**, *18* (3), 216–220. https://doi.org/10.1515/aut-2017-0022.

(18) Akter, M.; Khan, M. R. The Effect of Stitch Types and Sewing Thread Types on Seam Strength for Cotton Apparel. *International Journal of Scientific & Engineering Research* **2015**, *6* (7).

(19) Wang, L.; Chan, L. K.; Hu, X. Influence of Stitch Density to Stitches Properties of Knitted Products. *Research Journal of Textile and Apparel* **2001**, *5* (2), 46–53. https://doi.org/10.1108/RJTA-05-02-2001-B005.

(20) Admassu, Y.; Edae, A.; Getahun, G.; Senthil Kumar, S.; Saravanan, K.; Sakthivel, S. Experimental Analysis on the Effect of Fabric Structures and Seam Performance Characteristics of Weft Knitted Cotton Apparels. *Journal of Engineered Fibers and Fabrics* **2022**, *17*. https://doi.org/10.1177/15589250221113479.

(21) Midha, V. K.; Mukhopadhyay, A.; Chatopadhyay, R.; Kothari, V. K. Studies on the Changes in Tensile Properties of Sewing Thread at Different Sewing Stages. *Textile Research Journal* **2009**, *79* (13), 1155–1167. https://doi.org/10.1177/0040517508101456.

(22) Midha, V. K.; Mukhopadhyay, A.; Chattopadhyay, R.; Kothari, V. K. Effect of Process and Machine Parameters on Changes in Tensile Properties of Threads during High-Speed Industrial Sewing. *Textile Research Journal* **2010**, *80* (6), 491–507. https://doi.org/10.1177/0040517509338311.

(23) Sundaresan, G.; Hari, P. K.; Salhotra, K. R. Strength Reduction of Sewing Threads during High Speed Sewing in an Industrial Lockstitch Machine: Part I - Mechanism of Thread Strength Reduction. *International Journal of Clothing Science and Technology* **1997**, *9* (5), 334–345. https://doi.org/10.1108/09556229710185460.

(24) Sundaresan, G.; Salhotra, K. R.; Hari, P. K. Strength Reduction in Sewing Threads during High Speed Sewing in Industrial Lockstitch Machine: Part II: Effect of Thread and Fabric Properties. *International Journal of Clothing Science and Technology* **1998**, *10* (1), 64–79. https://doi.org/10.1108/09556229810205303.

(25) Rengasamy, R. S.; Wesley, S. Tensile Behavior of Different Types of Sewing Threads Observed under Simple-Tensile, Loop and Knot Tests. *Journal of Textile and Apparel, Technology and Management* **2011**, *7* (2).

(26) Abrishami, S.; Ezazshahabi, N.; Mousazadegan, F. Analysis of the Stress Relaxation Behaviour of Sewing Threads in the Straight and Loop Form. *The Journal of The Textile Institute* **2021**, *112* (4), 596–609. https://doi.org/10.1080/00405000.2020.1768773.

(27) Mouritz, A. P.; Cox, B. N. A Mechanistic Approach to the Properties of Stitched Laminates. *Composites Part A: Applied Science and Manufacturing* **2000**, *31* (1), 1–27. https://doi.org/10.1016/S1359-835X(99)00056-1.

(28) Mouritz, A. P.; Cox, B. N. A Mechanistic Interpretation of the Comparative In-Plane Mechanical Properties of 3D Woven, Stitched and Pinned Composites. *Composites Part A: Applied Science and Manufacturing* **2010**, *41* (6), 709–728.

https://doi.org/10.1016/j.compositesa.2010.02.001.

(29) Villanueva, R.; Ganta, D.; Guzman, C. Mechanical, in-Situ Electrical and Thermal Properties of Wearable Conductive Textile Yarn Coated with Polypyrrole/Carbon Black Composite. *Mater. Res. Express* **2018**, *6* (1), 016307. https://doi.org/10.1088/2053-1591/aae607.

(30) Ardalan, S.; Hosseinifard, M.; Vosough, M.; Golmohammadi, H. Towards Smart Personalized Perspiration Analysis: An IoT-Integrated Cellulose-Based Microfluidic Wearable Patch for Smartphone Fluorimetric Multi-Sensing of Sweat Biomarkers. *Biosensors and Bioelectronics* **2020**, *168*, 112450. https://doi.org/10.1016/j.bios.2020.112450.

(31) Qi, H. J.; Boyce, M. C. Constitutive Model for Stretch-Induced Softening of the Stress–Stretch Behavior of Elastomeric Materials. *Journal of the Mechanics and Physics of Solids* **2004**, *52* (10), 2187–2205. https://doi.org/10.1016/j.jmps.2004.04.008.

(32) Julio García Ruíz, M.; Yarime Suárez González, L. Comparison of Hyperelastic Material Models in the Analysis of Fabrics. *International Journal of Clothing Science and Technology* **2006**, *18* (5), 314–325. https://doi.org/10.1108/09556220610685249.

(33) Khiêm, V. N.; Krieger, H.; Itskov, M.; Gries, T.; Stapleton, S. E. An Averaging Based Hyperelastic Modeling and Experimental Analysis of Non-Crimp Fabrics. *International Journal of Solids and Structures* **2018**, *154*, 43–54. https://doi.org/10.1016/j.ijsolstr.2016.12.018.

(34) Gong, Y.; Peng, X.; Yao, Y.; Guo, Z. An Anisotropic Hyperelastic Constitutive Model for Thermoplastic Woven Composite Prepregs. *Composites Science and Technology* **2016**, *128*, 17–24. https://doi.org/10.1016/j.compscitech.2016.03.005.

(35) Peng, X.; Guo, Z.; Du, T.; Yu, W.-R. A Simple Anisotropic Hyperelastic Constitutive Model for Textile Fabrics with Application to Forming Simulation. *Composites Part B: Engineering* **2013**, *52*, 275–281. https://doi.org/10.1016/j.compositesb.2013.04.014.

(36) Peng, X. Q.; Guo, Z. Y.; Zia-Ur-Rehman; Harrison, P. A Simple Anisotropic Fiber Reinforced Hyperelastic Constitutive Model for Woven Composite Fabrics. *Int J Mater Form* **2010**, *3* (1), 723–726. https://doi.org/10.1007/s12289-010-0872-3.

(37) Geršak, J.; Knez, B. Reduction in Thread Strength as a Cause of Loading in the Sewing Process. *International Journal of Clothing Science and Technology* **1991**, *3* (4), 6–12. https://doi.org/10.1108/eb002978.